\documentclass[preprint2,longabstract]{aastex}

\usepackage{rotating}

\shorttitle{PAH in Galaxies at z $\sim$ 0.1}
\shortauthors{O'Dowd et al. 2009}

\begin{document}

\title{Polycyclic Aromatic Hydrocarbons in Galaxies at z $\sim$ 0.1: the Effect of Star Formation and AGN} 
\author{Matthew J. O'Dowd, David Schiminovich}
\affil{Astronomy Department, Columbia University, New York, NY 10027, USA}
\email{matt@astro.columbia.edu}
\author{Benjamin D. Johnson}
\affil{Institute of Astronomy, Madingley Road, Cambridge CB3 0HA, UK}
\author{Marie A. Treyer, Christopher D. Martin, Ted K. Wyder}
\affil{California Institute of Technology, MC 405-47, 1200 East California Boulevard, Pasadena, CA 91125, USA}
\author{S. Charlot}
\affil{Institut d'Astrophysique de Paris, UMR 7095, 98 bis Bvd Arago, 75014, Paris, France}
\author{Timothy M. Heckman}
\affil{Department of Physics and Astronomy, Johns Hopkins University, Homewood Campus, Baltimore, MD 21218, USA}
\author{Lucimara P. Martins}
\affil{Space Telescope Science Institute, 3700 San Martin Drive, Baltimore, MD 21218, USA}
\author{Mark Seibert}
\affil{Observatories of the Carnegie Institution of Washington, 813 Santa Barbara Street, Pasadena, CA 91101, USA}
\author{J. M. van der Hulst}
\affil{Kapteyn Astronomical Institute, University of Groningen, the Netherlands}

\begin{abstract}

We present the analysis of the Polycyclic Aromatic Hydrocarbon
(PAH) spectra of a sample of 92 typical star forming galaxies at $0.03
< z < 0.2$ observed with the {\em Spitzer} IRS. We compare the
relative strengths of PAH emission features with SDSS optical diagnostics to
probe the relationship between PAH grain properties and 
star formation and AGN activity. Short-to-long wavelength PAH ratios,
and in particular the 7.7~\micron-to-11.3~\micron~ feature ratio, are
strongly correlated with the star formation diagnostics $D_n(4000)$ and
H$\alpha$ equivalent width, increasing with younger stellar
populations. 
This ratio also shows a significant
difference between active and non-active galaxies, with the active
galaxies exhibiting weaker
7.7~\micron~ emission. A hard radiation field as measured
by $[OIII]/H\beta$ and $[NeIII]_{15.6~\mu m}/[NeII]_{12.8~\mu m}$
affects PAH ratios differently 
depending on whether this field results from starburst activity or an
AGN. Our results are consistent with a picture in
which larger PAH molecules grow more efficiently in richer media
and in which smaller PAH molecules are preferentially destroyed by AGN.

\end{abstract}

\keywords{ galaxies: active --- galaxies: ISM --- ISM: lines and bands
  --- ISM: molecules --- infrared: galaxies
  --- techniques: spectroscopic}

\section{INTRODUCTION}

The mid-infrared spectra of star forming galaxies are punctuated
by a series of broad peaks that dominate the emission between
3 and $19~\mu$m. These bands are generally accepted to result from
the vibrational modes of Polycyclic Aromatic Hydrocarbons (PAHs)
\citep{Leger, Allamandola}.
PAH molecules --- planar lattices of aromatic rings containing from
10's to 100's of carbon atoms --- suffuse the interstellar medium of
our own Galaxy. Their vibrational modes are excited by absorption of
UV photons, and in heavily star forming galaxies the PAH bands alone
can contribute a substantial fraction of the reprocessed light. 
Given that bolometric infrared (IR) emission has long been an important
measure of star formation due to the large fraction of
dust-absorbed starlight that is re-emitted in the IR, the PAH bands
themselves offer great promise as a more detailed diagnostic of star
formation history and as a probe of star formation in heavily
dust-obscured galaxies.

A major factor limiting the diagnostic use of galactic PAH emission
is that the detailed physics of large PAH molecules is poorly
understood. Models of stochastic heating of dust 
grains predict that the relative strengths of PAH bands are
dependent on the size distribution of PAH grains and on the ionization
state of the molecules \citep{Schutte, DraineLi}. However 
laboratory tests of PAHs have been limited to smaller size
molecules than dominate the interstellar medium \citep{Oomens, Kim}. 
As a result, the confidence with which the results of these models can
be applied to interstellar PAH spectra is still uncertain.
Nonetheless, such models indicate that measurements of relative PAH
band strengths will provide valuable measures of PAH
growth and destruction and of the ambient radiation field.

Observationally, the overall shape of the PAH spectrum shows
remarkable similarity across a broad range of environments within our
own Galaxy, and across a range of star formation histories in other
galaxies. However there are still significant variations observed in
the relative strengths of certain PAH emission 
bands. {\em Infrared Space Observatory} ({\em ISO}) and {\it Spitzer}
studies \citep{Peeters, Smith07, Galliano} have revealed trends
between PAH band ratios and various properties of the galaxy,
including AGN activity, star formation history, and/or galaxy
morphology.
Yet, because these properties are themselves related, it has been
difficult to disentangle primary physical processes
responsible for variations in PAH spectra.

To begin to understand the links between galaxies' PAH emission and
their physical properties, it is essential to study a sample spanning
the full range of normal galaxy properties, and this sample must
include extensive multi-wavelength data to enable characterization of
convolved physical properties.
These were the primary driving goals behind the Spitzer SDSS GALEX
Spectroscopic Survey (SSGSS). The SSGSS sample covers galaxies from
the blue cloud to the red sequence and transitional galaxies in
between, spanning two orders of magnitude in stellar mass, colour, and
dust attenuation. It combines both
broad-wavelength-coverage and high-resolution IRS spectroscopy with a
thorough suite of multi-wavelength data spanning the far ultraviolet
to the far infrared. This makes it the ideal data set for studying the
connection between galaxies' detailed IR emission and their physical
properties. 

In this paper we study the effect of star formation, metallicity, radiation field,
and AGN incidence on PAH molecules by looking at the connection
between relative PAH emission strengths and optical diagnostics of
these properties. In Section~\ref{obs} we describe the sample 
selection, data reduction and spectral decomposition. In 
Section~\ref{res} we study the relationships between the relative 
strengths of the strongest PAH bands and a range optical diagnostics. 
In Section~\ref{con} we present our conclusions.

\section{OBSERVATIONS AND SPECTRAL ANALYSIS}
\label{obs}

\begin{figure*}[ht!]
\centering
\includegraphics*[width=15cm]{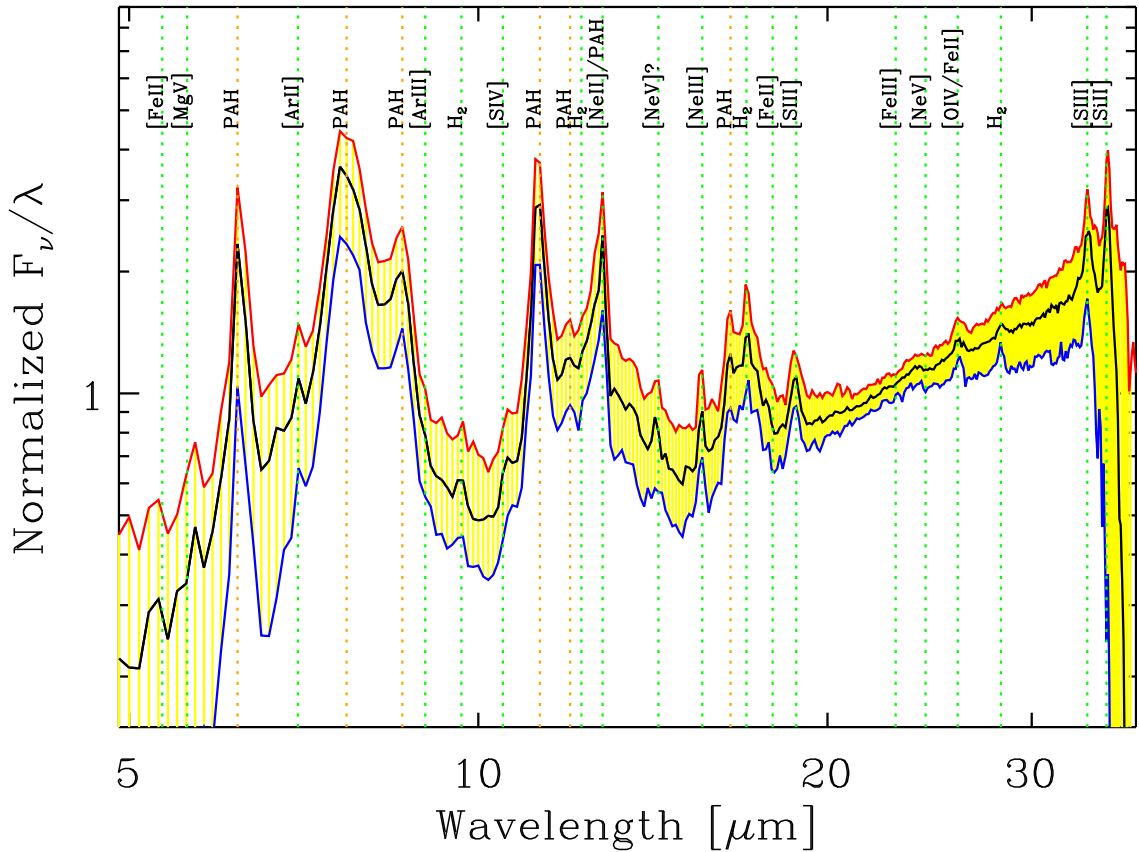}
\caption{Composite of the 92 SSGSS low-res spectra used in this paper
  (black line). Blue and red lines are the 16th and 84th percentile of
  the 2.5 sigma-clipped distribution at each wavelength. Spectra are
  normalized between 20~\micron~ and 24~\micron.
\label{compspec}}
\end{figure*}

\subsection{Sample, Observations and Data Reduction}
\label{sampobsred}

The full details of the SSGSS sample selection, observations, and data
reduction are reported in O'Dowd et al. (2009, in preparation). To summarize:

SSGSS is a Spitzer spectroscopic survey of 101 normal, star-forming
galaxies from the Lockman Hole region. This $\sim10$ square degree
field of low Galactic HI/cirrus is extensively surveyed in multiple 
wavebands. SSGSS galaxies were selected to have coverage by
SDSS, GALEX and IRAC (4 channel) and MIPS (24~\micron, 70~\micron).
We applied a surface brightness limit of I$_{5.8\mu m} > 0.75$~MJy~sr$^{-1}$, 
and a flux limit of F$_{24\mu m} > 1.5$~mJy. These criteria yielded 154
galaxies, from which we selected 101 galaxies to span the range
of normal galaxy properties by uniformly sampling $D_n(4000)$ vs. NUV-K space. 
This sample has a redshift range of $0.03 < z < 0.2$ with median
redshift $z_{med}=0.08$, and a total infrared luminosity range of 
$3.7\times10^9 L_\odot < L_{TIR} < 3.2\times10^{11} L_\odot$, with median
$3.9\times10^{10} L_\odot$.

Spectroscopy was obtained in staring mode using the 'low-res'
Short-Low (SL) and Long-Low (LL) 
IRS modules for the entire sample, providing 5.2---38.0~\micron~ coverage with
resolving power of $\sim$60 to 125. 'Hi-res' Short-high (SH) spectroscopy was
obtained for a subsample of the brightest 33 galaxies, providing
9.9---19.6~\micron~ coverage at $R \approx 600$. For the bright subsample,
exposure times were 8~min for the SL and LL modules, and 16~min for SH. For the
remaining sample, exposure times were 8~min for SL and 16~min for LL.
All data were taken during Spitzer IRS campaign 37.

We used the 2-D Spitzer data products processed by the Spitzer
Pipeline version S15.3.0, which performed standard IRS
calibration (ramp fitting, dark subtraction, droop, linearity
correction and distortion correction, flat fielding, masking and
interpolation, and wavelength calibration).
Sky subtraction was performed manually, with sky frames constructed
from the 2-D data frames, utilizing the shift in galaxy spectrum
position between orders to obtain clean sky regions. IRSCLEAN (v.1.9)
was used to clean bad and rogue pixels. SPICE was used to  
extract 1-D spectra, and these spectra were finally combined and
stitched manually by weighted mean.

Due to problems with some observations, the
sample studied in this paper consists of 92 galaxies observed with the
low-res modules and a subset of 32 observed at hi-res. 
In addition, for subset of sources spectrophotometric calibration
failed for the SL 2$^{nd}$ order module, resulting in fluxes 
a factor of $\sim 2$ lower than expected based on the SL 1$^{nd}$ order
module spectrum and IRAC photometry. We eliminate these sources from
analyses involving the affected spectral region, as described in
Section~\ref{bad6p2}.

Figure~\ref{compspec} shows the composite low-res spectrum for
this sample. PAH features and complexes at 6.2, 7.7, 8.6, 11.3, 12,
12.6, and 17~\micron~ are prominent, as are a number of atomic
emission lines. A thermal dust continuum component dominates redward of
$\sim 13$~\micron. These composite spectra are normalized between
20~\micron~ and 24~\micron.

\subsubsection{Aperture Effects}

The short- and long-low resolution IRS modules have respective slit widths of
$\sim$3\farcs7 and 10\farcs5, corresponding to physical
sizes of 2---12~kpc and 10---34~kpc over the redshift range of the
sample, with respective medians 5.5~kpc and 15.6~kpc.
The reduction process makes a flux correction for aperture effects 
assuming a point-like source. 
The maximum correction is at the intersection of the SL and LL
modules, where the slit-loss is $\sim$15\% and $\sim$45\%
respectively. 
Some error may be introduced due the fact that the different
apertures sample different regions of the galaxy. Fortunately, the PAH
features of primary interest are between 6.2 and 11.3~\micron, and so
all lie in the short-low regime, and are thus largely self-consistent. 

Aperture effects may also arise from the wavelength-dependent PSF
width. In extended sources, this may result in increased sampling of
central regions at the expense of extended regions with increasing
wavelength. We use IRAC and MIPS photometry in the 6, 8, 16 \&
24~\micron~ bands to explore this effect, and find that the difference
due to aperture effects in the 7.7~\micron-to-11.4~\micron~ PAH
ratio is on average less than 15\%, and we find no systematic bias in
this effect with any of the optical properties studied in this paper.

The SDSS fibre diameter is 3\arcsec, corresponding to a physical
1.6---9.7~kpc, with a median of 4.5~kpc, over the redshift range of
the sample. This is sufficiently close to the scale sampled by the 
SL slit that aperture effects with SDSS data
are expected to be minimal.

\subsection{Spectral Fitting and PAH Strengths} \label{fitting}

To measure the strengths of the PAH features we use the PAHFIT
spectral decomposition code of \cite{Smith07} (S07). This code
performs $\chi^2$ fitting of multiple spectral components, including
PAH features modeled as Drude profiles, the thermal dust continuum,
starlight, prominent emission lines, and dust extinction.
For ease of comparison with the results of S07, we use identical 
temperatures for the thermal continuum components and identical central
wavelengths and widths for the PAH features. 

We report a sample of the fitted fluxes of the most prominent PAH features and 
PAH feature complexes in Table~\ref{PAHstrengths}. These include the
6.2 and 8.3~\micron~ bands, which are discrete PAH features, and the
7.7, 11.3, and 17~\micron~ complexes, which are blends of 3, 2, and 4
subfeatures respectively.
Henceforth, we refer to individual PAH features and
complexes of multiple features simply as 'features'.
In Table~\ref{PAHstrengths} also report the fitted dust extinction
optical depth, $\tau_{9.7}$ and the integrated line strengths for
[NeII]$_{12.8~\mu m}$ and [NeIII]$_{15.6~\mu m}$. The full table,
including the entire data set and uncertainties for all parameters, is
available in the {\em Astrophysical Journal} electronic edition.

\begin{figure*}[ht!]
\centering
\centering
\includegraphics*[width=14cm]{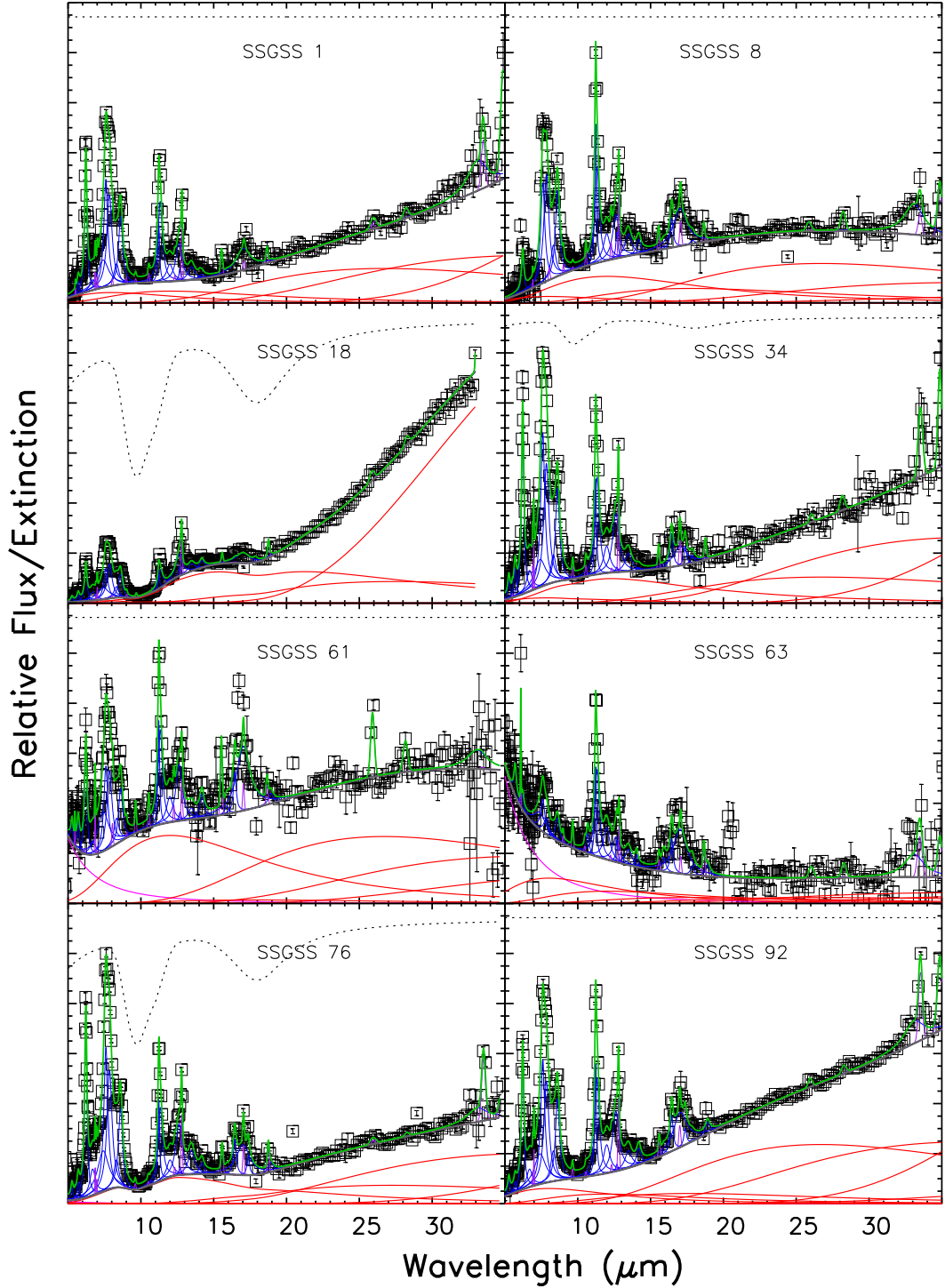}
\caption{Examples of PAHFIT decompositions of SSGSS spectra (black
  squares) in relative $\nu F_\nu$, showing a range of spectral shapes. The best fit (green)
  is composed of thermal dust continua at set temperatures (red), PAH
  features (blue), stellar light (magenta), emission lines (purple),
  and silicate absorption (dotted). SSGSS catalog number is also given.
\label{pahfitplots}}
\end{figure*}

Figure~\ref{pahfitplots} shows examples of PAHFIT decompositions of the SSGSS
spectra. 

\begin{deluxetable}{cccccccccccc}
\tabletypesize{\tiny}
\tablewidth{0pt}
\tablecaption{
\label{PAHstrengths}}
\tablehead{
\colhead{SSGSS} & \colhead{RA} & \colhead{DEC} & \colhead{z} & \colhead{$F_{6.2}$} & \colhead{$F_{7.7}$} &\colhead{$F_{8.6}$} & \colhead{$F_{11.3}$} & \colhead{$F_{17}$} & \colhead{$F_{[NeII]}$} & \colhead{$F_{[NeIII]}$} &\colhead{$\tau_{9.7}$}
}
\startdata
 1 & 160.34398 & 58.89201 & 0.066 & 1.35E-16 & 4.35E-16 & 8.89E-17 & 1.09E-16 & 4.72E-17 & 1.08E-17 & 1.17E-18&   ...   \\
 2 & 159.86748 & 58.79165 & 0.045 & 7.04E-17 & 2.14E-16 & 4.88E-17 & 5.43E-17 & 4.21E-17 & 3.51E-18 & 4.58E-18&   ...   \\
11 & 162.41000 & 59.58426 & 0.047 & 4.56E-17 & 1.58E-16 & 3.32E-17 & 5.37E-17 & 3.40E-17 & 2.67E-18 & 2.02E-19&   ...   \\
12 & 162.36443 & 59.54812 & 0.072 & 1.55E-16 & 6.61E-16 & 1.37E-16 & 1.99E-16 & 1.40E-16 & 3.41E-17 & 3.92E-18& 1.47\\
14 & 162.52991 & 59.54828 & 0.153 & 8.67E-17 & 3.11E-16 & 6.74E-17 & 8.93E-17 & 4.97E-17 & 1.06E-17 & 1.93E-18&   ...   \\
15 & 161.78737 & 59.63707 & 0.153 & 1.60E-17 & 5.81E-17 & 2.67E-17 & 3.88E-17 & 1.41E-17 & 8.37E-18 &   ...   &   ...   \\
16 & 161.48123 & 59.15443 & 0.072 & 8.38E-17 & 3.10E-16 & 6.03E-17 & 8.49E-17 & 6.12E-17 & 1.35E-17 & 1.42E-18&   ...   \\
17 & 161.59111 & 59.73368 & 0.047 & 5.68E-16 & 2.02E-09 & 4.11E-16 & 4.74E-16 & 2.81E-16 & 4.71E-17 & 1.11E-17& 0.94\\
27 & 161.11412 & 59.74155 & 0.072 & 1.32E-16 & 1.19E-09 & 2.67E-16 & 3.29E-16 & 1.97E-16 & 6.58E-17 & 1.38E-17&   ...   \\
28 & 161.71980 & 56.25187 & 0.103 & 1.98E-18 & 1.95E-16 & 4.68E-17 & 8.54E-17 & 2.67E-17 & 7.68E-18 & 4.23E-18&   ...   \\
30 & 162.26756 & 56.22390 & 0.046 & 1.13E-16 & 4.02E-16 & 7.74E-17 & 1.00E-16 & 3.01E-17 & 5.10E-18 & 4.63E-18&   ...   \\
32 & 163.00845 & 56.55043 & 0.117 & 7.61E-17 & 2.64E-16 & 5.72E-17 & 6.91E-17 & 1.75E-17 & 9.13E-18 & 4.76E-18&   ...   \\
33 & 161.92709 & 56.31395 & 0.185 & 2.99E-17 & 1.61E-16 & 2.68E-17 & 5.48E-17 & 3.89E-17 & 7.02E-18 & 2.23E-18& 0.66\\
34 & 161.75783 & 56.30670 & 0.046 & 9.03E-17 & 2.76E-16 & 6.15E-17 & 7.29E-17 & 3.47E-17 & 1.01E-17 & 3.14E-18& 0.20\\
39 & 162.04231 & 56.38041 & 0.074 & 5.47E-17 & 1.97E-16 & 4.41E-17 & 5.54E-17 & 1.49E-17 & 8.42E-18 & 1.47E-18&   ...   \\
45 & 161.76901 & 56.34029 & 0.113 & 2.41E-17 & 2.01E-16 & 4.64E-17 & 6.85E-17 & 3.07E-17 & 1.73E-17 & 1.02E-18&   ...   \\
47 & 163.39658 & 56.74202 & 0.102 & 7.47E-17 & 2.48E-16 & 5.14E-17 & 6.64E-17 & 2.89E-17 & 1.46E-17 & 2.18E-18&   ...   \\
48 & 163.44330 & 56.73859 & 0.200 & 7.44E-17 & 2.92E-16 & 5.86E-17 & 7.52E-17 & 1.88E-17 & 7.51E-18 & 1.98E-18&   ...   \\
54 & 163.26968 & 56.55812 & 0.115 & 2.12E-16 & 7.50E-16 & 1.54E-16 & 1.96E-16 & 1.30E-16 & 1.41E-17 & 1.62E-18& 0.82\\
61 & 163.19810 & 56.48840 & 0.073 & 3.14E-17 & 1.23E-16 & 2.00E-17 & 4.14E-17 & 4.56E-17 & 3.36E-17 & 3.98E-18&   ...   \\
62 & 163.09050 & 56.50836 & 0.133 & 1.03E-16 & 3.82E-16 & 7.03E-17 & 9.18E-17 & 5.97E-17 & 2.15E-17 &   ...   &   ...   \\
64 & 163.53931 & 56.82104 & 0.073 & 2.62E-16 & 9.32E-16 & 1.85E-16 & 2.04E-16 & 1.69E-16 & 4.45E-17 & 6.49E-18& 1.14\\
65 & 158.22482 & 58.10917 & 0.118 & 1.82E-16 & 6.69E-16 & 1.26E-16 & 1.74E-16 & 1.00E-16 & 2.72E-17 & 4.82E-18& 0.29\\
69 & 159.04880 & 57.72258 & 0.076 & 8.83E-17 & 2.46E-16 & 5.92E-17 & 7.23E-17 & 4.54E-17 & 1.38E-17 & 1.91E-18&   ...   \\
70 & 159.34668 & 57.52069 & 0.090 & 4.72E-17 & 2.17E-16 & 4.75E-17 & 6.48E-17 & 2.57E-17 & 8.37E-18 & 2.46E-18&   ...   \\
73 & 158.91122 & 57.59536 & 0.080 & 1.88E-17 & 9.81E-17 & 2.06E-17 & 2.47E-17 & 2.50E-17 & 7.12E-18 &   ...   &   ...   \\
77 & 158.91344 & 57.71219 & 0.044 & 3.93E-17 & 1.62E-16 & 3.22E-17 & 3.56E-17 & 1.60E-17 & 1.12E-17 & 5.46E-18&   ...   \\
83 & 159.73558 & 57.26361 & 0.119 & 3.13E-19 & 3.06E-16 & 6.76E-17 & 8.36E-17 & 4.79E-17 & 4.69E-18 & 3.78E-18&   ...   \\
94 & 159.63510 & 57.40035 & 0.061 & 7.79E-17 & 2.77E-16 & 5.33E-17 & 6.52E-17 & 4.08E-17 & 1.44E-17 & 5.61E-19&   ...   \\
95 & 161.48724 & 57.45520 & 0.115 & 7.48E-17 & 2.66E-16 & 5.84E-17 & 8.49E-17 & 2.52E-17 & 1.33E-17 & 1.04E-19& 0.79\\
98 & 160.29099 & 56.93161 & 0.050 & 6.40E-17 & 7.98E-16 & 2.18E-16 & 2.64E-16 & 1.37E-16 & 1.81E-17 & 1.00E-17&   ...   \\
99 & 160.30701 & 57.08246 & 0.077 & 1.62E-16 & 5.30E-16 & 1.11E-16 & 1.23E-16 & 8.50E-17 & 2.09E-17 & 1.41E-17&   ...   \\
 & \enddata
\begin{small}
  Integrated fluxes in W~m$^{-2}$ for PAH features and
  complexes and for $[NeII]_{12.8~\mu m}$ and $[NeIII]_{15.6~\mu m}$ 
  emission lines derived from PAHFIT spectral decomposition of the
  subsample of SSGSS galaxies with hi-res spectroscopy.  Also reported
  are the redshift of the galaxy, z, and dust extinction given as
  $\tau_{9.7}$, the optical depth at 9.7~\micron. PAH strengths and
  absorption come from fitting of low-res data while emission lines
  come from fitting of hi-res data. Results for the full
  data set, including uncertainties for all values, is included in the
  electronic edition of the {\em Astrophysical Journal}.
\end{small}
\end{deluxetable}

\subsubsection{Dust Extinction}

Dust extinction can have a pronounced effect on PAH feature strengths 
due to the strong silicate absorption features at $\sim$10 and
$\sim$18~\micron. PAHFIT provides the option to fit a fixed extinction
template as a free parameter. The standard template, which we utilize,
models absorbing dust mixed with the emitting stars and grains, as
opposed to an absorbing screen.
Introducing this free parameter has the potential to lead to
degenerate solutions and hence to large uncertainties in PAH feature
strengths, and so we investigated the effect of inclusion/omission of
extinction fitting on the fit results. We find that the strengths measured 
for the 6.2~\micron~ and 7.7~\micron~ PAH features are very similar
whether or not dust extinction is included, with average
deviations of less than 5\% and no systematic bias. On the other
hand, the 8.6~\micron~ and 11.3~\micron~ feature strengths appear to be
systematically overestimated without the inclusion of silicate
absorption by an average of 40\% and 20\% respectively.

Due to this systematic effect, the results that we present in this
paper utilize PAHFIT fits including extinction fitting. 33\% of
galaxies provide better fits with some level of silicate absorption,
with a median fitted extinction of $\tau = 0.7$.
Having repeated all analyses excluding silicate absorption we
find that our primary results are essentially unchanged. 
We make only minimal use of the strongly affected 8.6~\micron~ feature. 

\subsubsection{Spline Fitting}

A commonly used alternative method for measuring PAH strengths 
subtracts a fitted spline to represent the continuum and integrate the
residuals within each PAH band.
This method has the advantage of being independent
of assumptions about the physical contributions to the fitted
spectrum. 
However, having applied this method to all spectra, we find that it
significantly underestimates PAH strengths because it
misses the PAH features' extended wings, and because the continuum is
typically overestimated as the spline fitting points usually include
substantial PAH flux. Worse, we find that PAH ratios are particularly
sensitive to the systematic errors of this method as it depends
strongly on the variable widths and level of blending of
PAH features. As a result, we only report results based on the PAHFIT
decompositions. 

\subsubsection{Problematic 6.2~\micron~ Feature Fits}
\label{bad6p2}

The lower sensitivity of the 2$^{nd}$ order of the IRS SL
module results in low signal for a number of sources, and hence to
problems in fitting the 6.2~\micron~ feature in some cases. 
Although some 6.2~\micron~ PAH flux is clearly present in almost all
galaxies, PAHFIT failed to find a good fit to this feature in 18
cases. In many of these cases, the fit significantly underestimated
the 6.2~\micron~ flux, perhaps because of a small mismatch between the 
width or peak position of the feature and the fitted Drude profile.

It is possible to estimate PAH fluxes independently of PAHFIT, however
for the analysis of PAH ratios we restrict ourselves to self-consistant
PAH flux measurements using PAHFIT. Hence, we omit
failed 6.2~\micron~ fits from analyses that directly involve this feature.
By BPT designation (see Sect.~\ref{BPTdiag}) 
these include 10 non-active galaxies, 3 composite sources, and 5 AGN. 

As mentioned in Section~\ref{sampobsred}, a calibration problem with
the SL 2$^{nd}$ order resulted in aberrantly low fluxes in this
order for a further subset of the galaxies. We identify 14 galaxies whose
IRAC flux in the 5.7~\micron~ band is significantly greater than the
SL 2$^{nd}$ order flux integrated across this band. We omit these
galaxies from analyses of the 6.2~\micron~ feature. These include 10
non-active galaxies and 4 composite sources.

In total, 30 galaxies are omitted from analyses and plots involving
the 6.2~\micron~ feature. 
While this introduces unquantifiable biases, we still include such analyses
where we feel that they remain informative. 
However the focus of this work is on the more robust PAH measurements
- in particular th 7.7~\micron~ and 11.3~\micron~ features.

\subsubsection{The Relative Dominance of Star-Formation and AGN}
\label{BPTdiag}

\begin{figure}[ht!]
\epsscale{1.0}
\includegraphics*[width=7cm]{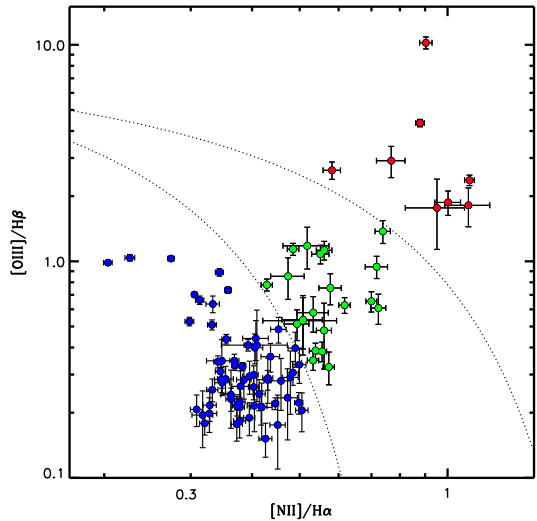}
\caption{BPT plot \citep{Baldwin} for the SSGSS sample, coloured coded
according to BPT designation; blue points are galaxies dominated by
star formation, green points are galaxies with significant stellar and
AGN components, and red points are AGN-dominated. Strictly, these
designations are measures of the relative dominance of AGN and star-forming
components within the observed aperture. However, for convenience, we utilize
the standard naming convention and refer to these as star-forming
galaxies, composite sources, and AGN, respectively. The upper and 
lower dotted lines represent the \citet{Kewley} and
\citet{Kauffmann03} divides respectively.
\label{BPT}}
\end{figure}

Throughout this paper we colour-code to indicate location in 
log~$[OIII]_{5007}/H\beta$ vs. log~$[NII]_{6583}/H\alpha$ space: the so-called BPT diagram of 
\cite{Baldwin} (Fig.~\ref{BPT}). Red points indicate more powerful AGN satisfying the
\cite{Kewley} designation, green points indicate composite sources
with weaker active components between the Kewley and the
\cite{Kauffmann03} designation, while black points indicate galaxies
dominated by stellar light, below the Kauffmann designation. 

BPT designation must be interpreted carefully. Rather than an absolute
discriminator between star-forming galaxies, composite sources, and
AGN, for this study we instead interpret it as a measure of the relative
dominance of star-formation and AGN with the Sloan aperture. The
similarity of the Sloan and {\it Spitzer} SL apertures means that BPT
designation measures this relative dominance within roughly the same galactic
region responsible for the observed PAH emission. The regions probed
in this study are larger than those probed in earlier, local
studies. This is an important difference as it means that we observe a
more global stellar population, where any AGN component is less overwhelming.

\subsubsection{Correlation of PAH Ratios and Comparison to Other  Studies}
\label{otherstudies}

\begin{figure*}[ht!]
\hspace{-3cm}
\centering
\includegraphics[width=13cm]{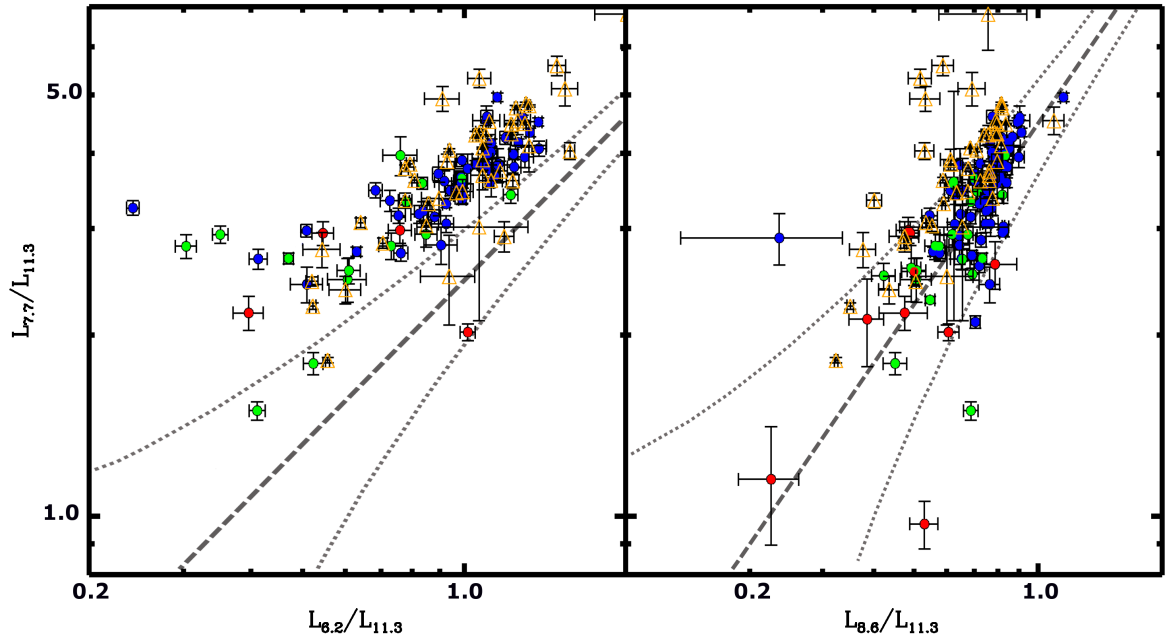}
\caption{
The PAH band ratios $L_{6.2}/L_{11.3}$ versus $L_{7.7}/L_{11.3}$
(left) and $L_{8.6}/L_{11.3}$ versus $L_{7.7}/L_{11.3}$ (right).  
Yellow triangles are SINGS galaxies from \citet{Smith07}, derived
using PAHFIT with the same parameters. The dashed and dotted lines
show the best linear regression and the dispersion for the PAH ratios
of Galactic HII regions and dwarf spiral and starburst galaxies found
by \citet{Galliano}, fit using a different technique.
Colour-coding is the same as in Figure~\ref{BPT}. 
\label{vsGalliano}}
\end{figure*}

Ratios of the integrated fluxes of the fitted Drude profiles were used
to determine PAH luminosity ratios, designated 
$L_{\lambda 1}/L_{\lambda 2}$, where $\lambda$ is the
central wavelength of the feature in micrometers.
Figure~\ref{vsGalliano} shows the tight correlations between different
short-to-long wavelength PAH ratios. 
$L_{6.2}/L_{11.3}$ versus $L_{7.7}/L_{11.3}$ 
follows a tight locus with $\sim$0.25~dex scatter. 
$L_{8.6}/L_{11.3}$ versus $L_{7.7}/L_{11.3}$ shows a
similar scatter, but broadens towards low ratio values. The tight
correlations suggest that lower-wavelength PAH features are strongly
coupled, and are less-strongly tied to the 11.3~\micron~ feature.

In this figure,
we also compare our results to two other spectroscopic studies of PAH
intensities:

S07 present spectral decomposition of IRS observations of 59 SINGS
galaxies, and we use an identical spectral decomposition
technique (see Sect.~\ref{fitting}). 
An important difference between S07 and our survey is, with their
local targets and shorter exposure times, they 
are dominated by the highest surface brightness regions near
the nucleus, while our observations sample more of the extended structure.
Their reported PAH ratios are very similar to ours, with the exception
that the S07 sample have, on average, higher $L_{7.7}/L_{11.3}$, and
lack the small number of galaxies with $L_{6.2}/L_{11.3} < 0.3$ that we observe. 

\cite{Galliano} (G08) present spectral decomposition of ISO and
Spitzer IRS observations of Galactic HII regions and dwarf
spiral and starburst galaxies. We take the results of their most
comparable decomposition technique, in which fit they Lorenzian
profiles to the PAH bands (as opposed to the Drude profiles used by PAHFIT). While
their sample show very similar trends in the 7.7~\micron~ and 8.6~\micron~
features versus the 11.3~\micron~ feature, they find that the 6.2~\micron~
feature is on average stronger by a factor of 1.3 to 2 than both our
results and those of S07. This discrepancy is likely due to the
differences in the fitting techniques.

\section{RESULTS AND DISCUSSION}
\label{res}

\subsection{PAH Ratios as Measures of Grain Size and Ionization}
\label{sizeion}

Models of the stochastic heating of dust grains \citep{Tielens,
  DraineLi, Schutte} show
that the relative power emitted in a given PAH band is strongly
dependent on the distribution of grain sizes. 
In general, smaller dust grains emit more power in the shorter
wavelength bands and larger grains 
dominate longer wavelengths bands. In the regime of grain sizes expected in
interstellar dust (10's to 100's of carbon atoms), rapid drops are 
expected in the 6.2~\micron~ and $7.7~\mu$m bands relative to
longer-wavelength features, and in 6.2~\micron~ relative to 7.7~\micron,
with increasing grain size.

It is also expected that the ionization state of a PAH molecule will have a
dramatic effect on its spectrum \citep{DraineLi}.  
In particular, carbon-carbon vibrational modes are known to be
significantly more 
intense in ionized PAH molecules \citep{Tielens}. 
As the 6.2~\micron~ and $7.7~\mu$m bands result from radiative 
relaxation of CC stretching modes, the ratios of these bands
to those arising from carbon-hydrogen modes such as the 11.3~\micron~
feature are expected to drop by an order of magnitude between 
completely neutral and completely ionized PAH clouds. At the same
time, the 6.2~\micron~ feature should change little relative to the
$7.7~\mu$m  feature as the ionization fraction changes.

\begin{figure*}[ht!]
\hspace{-3cm}
\centering
\includegraphics[width=13cm]{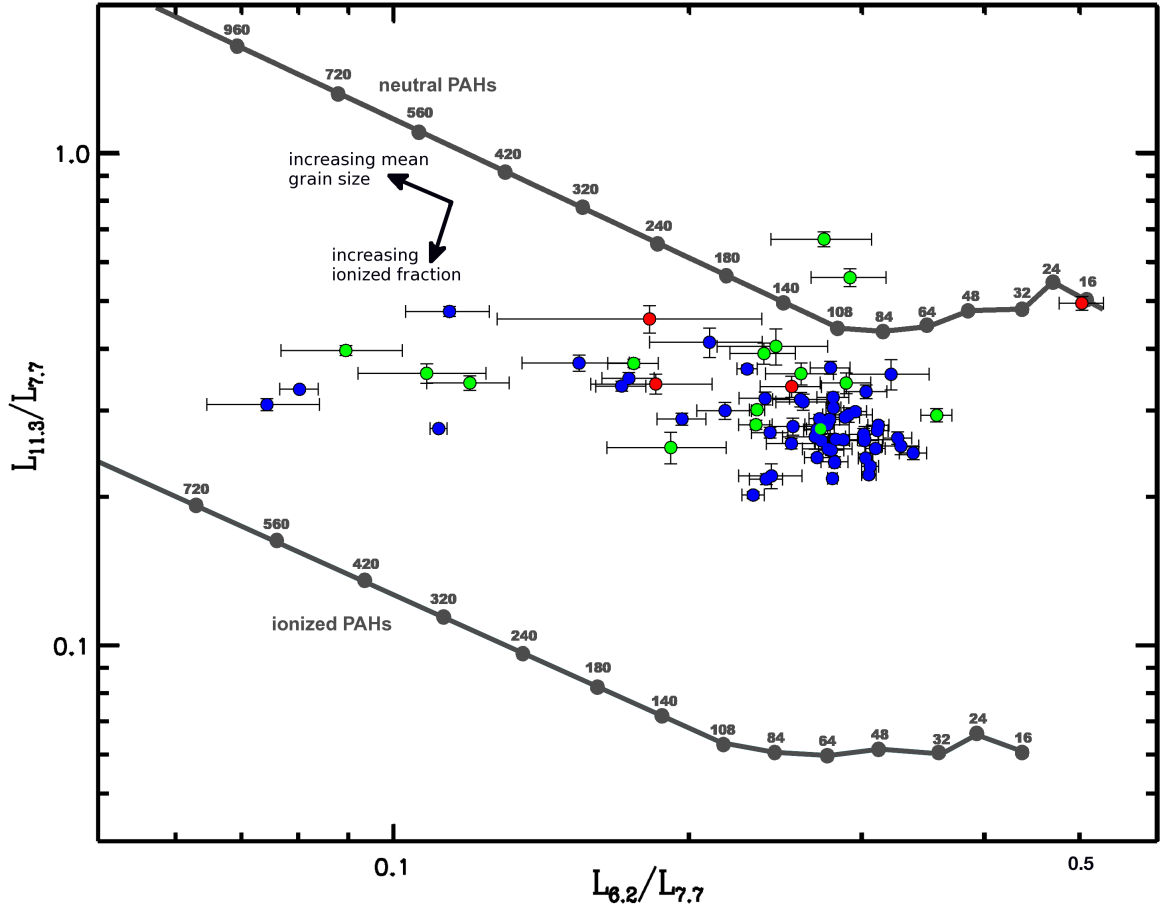}
\caption{
The PAH band ratios $L_{6.2}/L_{7.7}$ versus $L_{11.3}/L_{7.7}$.
Grey lines show the expected PAH ratios as a function of grain size
and ionization state from the models of \cite{DraineLi}. The lines
represent the expected ratios for fully neutral or fully ionized PAH
molecules of a given number of carbon atoms. These are meant to be
illustrative only, as real PAH ensembles will
contain mixtures of ionization states and grain sizes.
Colour-coding is the same as in Figure~\ref{BPT}. 
\label{DraineLi}}
\end{figure*}

Comparison of the PAH band power ratios $L_{6.2}/L_{7.7}$ to 
$L_{11.3}/L_{7.7}$ is useful in extricating these two effects,
with the former being sensitive to PAH grain size but relatively
unaffected by ionization as both bands result from the similar CC
modes, while the latter is very sensitive to ionization, and somewhat
less so to grain size.
Figure~\ref{DraineLi} shows these ratios
for our sample plotted over the \cite{DraineLi} models. 
Most of the sample is restricted to a tight locus between 0.2 and 0.4 
in both $L_{6.2}/L_{7.7}$ and
$L_{11.3}/L_{7.7}$, although within this locus there is
a weak trend in the direction expected for constant ionization fraction
with changing grain size, albeit with significant scatter.
We discuss the distribution of active galaxies in this plot in Section~\ref{AGNSF}

\subsection{Comparison of PAH Ratios with Optical Diagnostics}
\label{opticalcorrelation}

\begin{figure*}[ht!]
\centering
\includegraphics[angle=90,width=16cm]{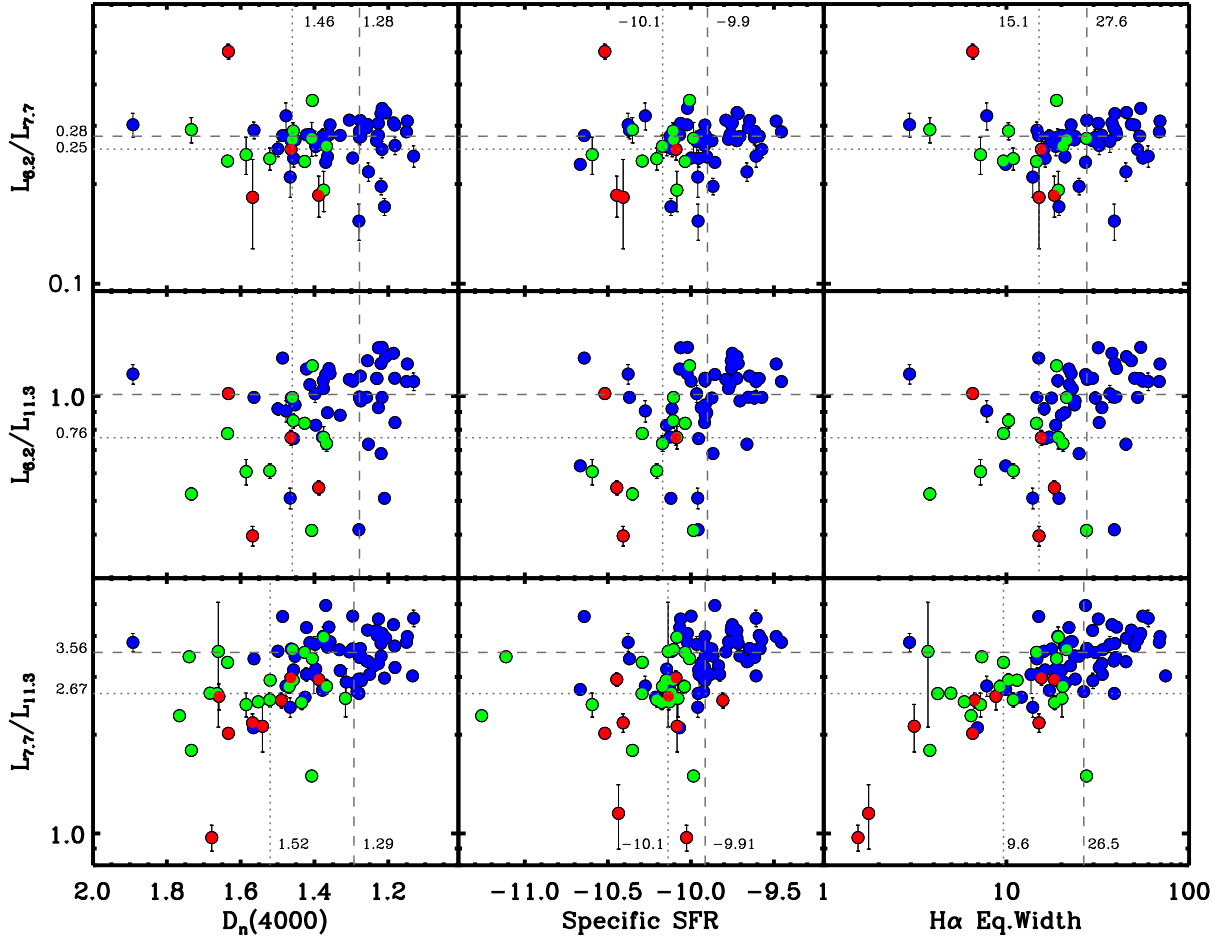}
\caption{PAH ratio versus $D_n(4000)$, SSFR, and H$\alpha$ equivalent
  width. Colour-coding is the same as in Figure~\ref{BPT}. The
  lines show the median values for galaxies with no AGN component
  (dashed) and AGN/composite sources (dotted).
\label{PAHratiovsSF}}
\end{figure*}

We compare the PAH spectra resulting from our PAHFIT models to SDSS
optical diagnostics of star formation history and AGN activity. 
Figure~\ref{PAHratiovsSF} 
shows the ratio of short-to-long wavelength PAH bands 
($L_{6.2}/L_{7.7}$, $L_{6.2}/L_{11.3}$, and
$L_{7.7}/L_{11.3}$) versus the star formation diagnostics
$D_n(4000)$, Specific Star Formation Rate (SSFR), and H$\alpha$
Equivalent Width (H$\alpha$~EW). $D_n(4000)$
\citep{KauffmannSDSS2003} measures the 4000\AA~ break over a narrower
bandpass than the standard $D(4000)$ \citep{Bruzual1983}, and provides
a measure of the luminosity-weighted stellar age.  
SSFR \citep{Brinchmann04} is the star formation rate as determined by
population synthesis models per unit stellar mass. 
The H$\alpha$ line provides a direct measure of recent star formation,
and in particular the presence of short-lived OB stars.

These star formation diagnostics are relatively free of the effects of
dust extinction; $D_n(4000)$ and H$\alpha$ EW because of the narrow
bandpasses over which these diagnostics are calculated, and SSFR
because it has been corrected for extinction \citep{Brinchmann04}.

The most striking correlations are seen relative to 
$L_{7.7}/L_{11.3}$, with the short wavelength
PAH band becoming more dominant with increasing recent star formation.
The ratios of the 6.2~\micron~ feature to longer wavelength features
follow similar, albeit weaker trends to $L_{7.7}/L_{11.3}$. 

Table~\ref{regressionresults} shows the results of the regression
analyses of PAH ratios to optical properties. We use Kendall's 
correlation to determine the significance of the trends, given as
probabilities of obtaining the observed data given the null hypothesis
of no correlation. We obtain a significance of $<$5\% 
for $L_{6.2}/L_{11.3}$ and $L_{7.7}/L_{11.3}$
versus all star formation diagnostics, while
$L_{6.2}/L_{7.7}$ shows no significant correlations.

\begin{figure*}[ht!]
\centering
\includegraphics[angle=90,width=16cm]{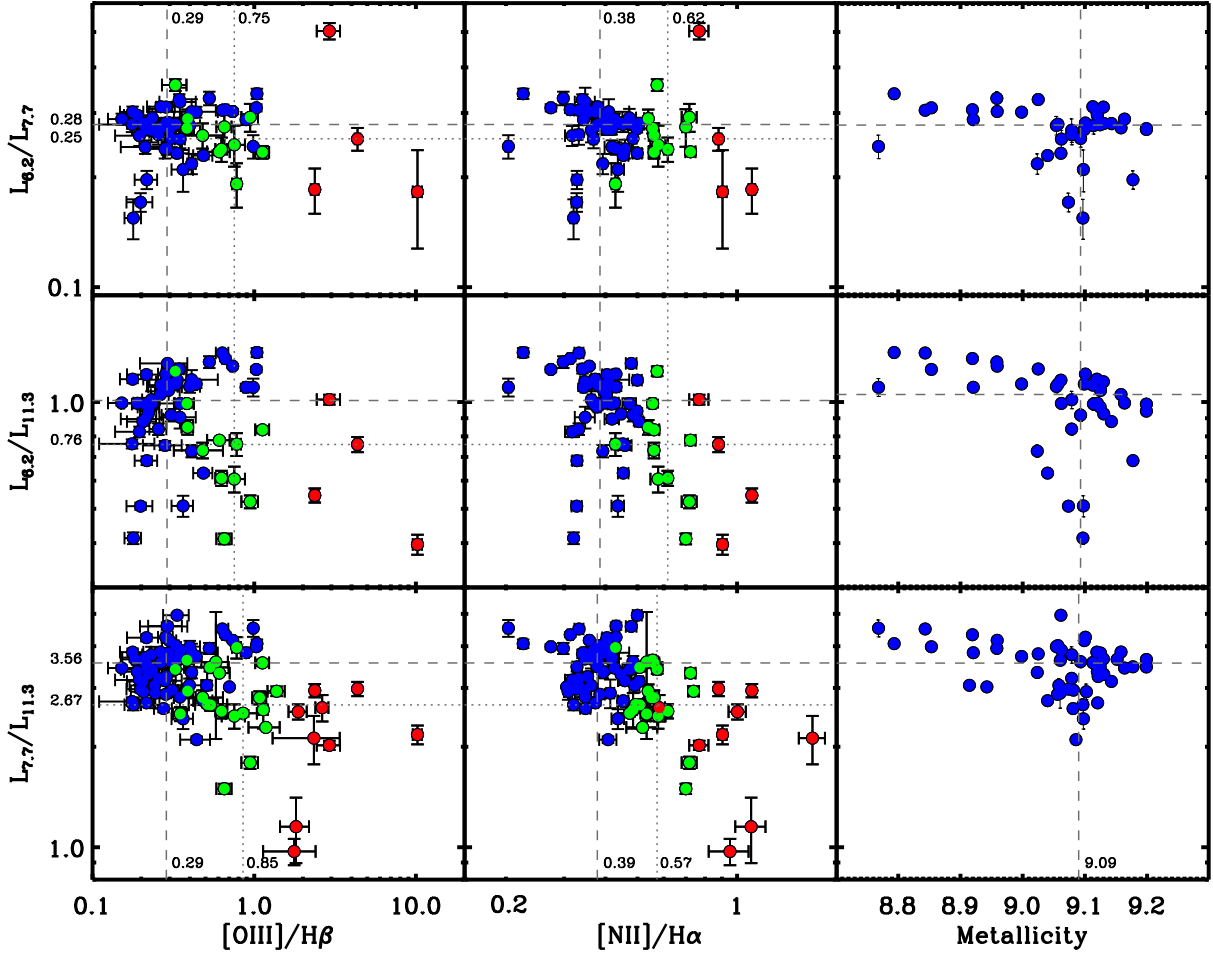}
\caption{PAH ratio versus $[NII]/H\alpha$ and $[OIII]/H\beta$. 
Colour-coding is the same as in Figure~\ref{BPT}.  The
  lines show the median values for galaxies with no AGN component
  (dashed) and AGN/composite sources (dotted).
\label{PAHratiovsIonization}}
\end{figure*}

$[OIII]/H\beta$ provides a measure of the ionization parameter
weighted by SFR, and hence of the hardness of the radiation
field present in the galaxy \citep{Kewley}. From
Figure~\ref{PAHratiovsIonization} (left) it 
can be seen that there is a global trend between this
line ratio and $L_{7.7}/L_{11.3}$, and no strong trends with
ratios involving the 6.2~\micron~ feature. 

The emission line ratio $[NII]/H\alpha$ provides a diagnostic of
gas-phase metallicity, although it is also affected by radiation field
hardness. This ratio strongly correlates with the age of the
stellar population. As expected, $[NII]/H\alpha$ follows the trends
observed with star formation diagnostics
(Fig.~\ref{PAHratiovsIonization}, middle). The best global correlation
is again with $L_{7.7}/L_{11.3}$, with the ratios of
short-to-long wavelength PAH bands decreasing with increasing
emission line ratio.

Table~\ref{regressionresults} also shows the results of the
correlation analysis for PAH ratios versus emission line ratios.

A more reliable measure of gas-phase metallicity, derived from 
a range SDSS optical nebular lines \citep{Tremonti} is available for a
subset of the sample, although contain no BPT-designated AGN or composite
sources (Fig.~\ref{PAHratiovsIonization}, right). 
The correlations of PAH ratios with this
metallicity measure are slightly weaker than that for $[NII]/H\alpha$,
suggesting that the trends observed in $[NII]/H\alpha$ may be
dominated by the trend with radiation hardness. However there is a
notable locus of high short-to-long wavelength PAH ratios for lower
metallicities. Galaxies in these upper left loci have the lowest 
metallicities ($12 + log(O/H) < 9$) and are among the youngest in the
sample ($D_n(4000) < 1.3$), and exhibit the hardest radiation fields
of the star-forming galaxies ($[OIII]/H\beta > 0.4$). 

It is known that PAH emission around 8~\micron~ is 
stronger in higher metallicity galaxies \citep{Engelbracht}. 
The dominance of short wavelength PAH features in the lowest
metallicity galaxies as seen in Figure~\ref{PAHratiovsIonization}
(right) may result from a difficulty in building large
PAH molecules in an under-rich medium. It may also result from other
age-dependent effects given the close link between metallicity and
age. However it is not expected to arise through the link between 
metallicity and radiation field hardness;
a hard radiation field will either
preferentially destroy small PAH molecules, decreasing this ratio, or
ionize molecules, which should not produce the observed change in 
$L_{6.2}/L_{7.7}$ with metallicity, as seen in Section~\ref{sizeion}. 

\begin{table*}[ht!]
\begin{center}
\begin{tabular}{llccc}
&&&&\\[-2ex]
Diagnostic & Sample & \Large{$\frac{L_{6.2}}{L_{7.7}}$} & \Large{$\frac{L_{6.2}}{L_{11.3}}$} & \Large{$\frac{L_{7.7}}{L_{11.3}}$}\\ [1ex]
\tableline
\tableline
           & full sample & ~~~~...~~~~ & -0.07 (1.6\%) & -0.26 ($<$0.01\%)\\
$D_n(4000)$&star-forming       & ... & ... & -0.39 (3\%)\\
           &AGN/composite       & ... & ... & ...\\
\tableline
    & full sample & ... & 0.2 (0.17\%) & 0.21 ($<$0.01\%)\\
SSFR&star-forming       & ... & ... & 0.13 (0.2\%)\\
    &AGN/composite       & ... & ... & ...\\
\tableline
             & full sample & ... & 0.16 (0.01\%) & 0.19 ($<$0.01\%)\\
H$\alpha$ EW&star-forming       & ... & 0.1 (0.5\%) & 0.16 (0.01\%) \\
             &AGN/composite       & ... & ... & 0.19 (2\%)\\
\tableline
                                     & full sample & ... & -0.5 (0.1\%) & -0.53 ($<$0.01\%)\\
\Large{$\frac{[NII]}{H\alpha}$}&star-forming       & ... & ... & ...\\
                               &AGN/composite       & ... & ... & ...\\
\tableline
                                     & full sample & ... & ... & -0.20 (1.2\%)\\
\Large{$\frac{[OIII]}{H\beta}$}&star-forming       & ... & ... & -0.07 (4\%)\\
                               &AGN/composite       & ... & ... & -0.25 (5\%)\\
\tableline
\end{tabular}
\caption{
Results of linear fits and regression analysis for
PAH ratios versus SDSS optical properties. The numbers given are 
the slopes of the best linear fits, followed in brackets by the
significance of the trend according to Kendall's rank
correlation. Dashes indicate that the correlation is not significant
at the 5\% level or better.
\label{regressionresults}}
\end{center}
\end{table*}

\subsection{Separating the Effects of AGN Activity and Star Formation} \label{AGNSF}

From Figures \ref{PAHratiovsSF} and \ref{PAHratiovsIonization} it
can be seen that galaxies with dominant AGN components (those above
the Kauffmann divide on the BPT diagram) have, on average, weaker
short-to-long wavelength PAH bands.  
There is a strong link between AGN incidence and the star formation
history of a galaxy; most notably, AGN are not apparent in the most
strongly star-forming galaxies, because of swamping of diagnostic
lines, dust obscuration, or for reasons related to the life-cycle of
the AGN. The link between age and BPT designation is
apparent from Figure~\ref{PAHratiovsSF}. This makes it 
difficult to determine whether AGN activity, star formation history,
or a combination of the two are responsible for the trends observed
between PAH ratios and optical properties.

\subsubsection{The Effect of Star Formation}

We repeat the regression analysis and linear fits from
Section~\ref{opticalcorrelation} independently for subsamples of
star-forming galaxies and AGN/composite sources. For both subsamples, 
the correlations with $L_{7.7}/L_{11.3}$ still hold for
H$\alpha$~EW. However, for $D_n(4000)$ and SSFR the correlation
remains significant at better than the 5\% level only for the
star-forming subsample. For $L_{6.2}/L_{11.3}$ the
correlation only holds for the star formation-dominated subsample with
respect to H$\alpha$~EW. 

In general, it appears that star formation 
history does have an influence on ratio of the $7.7\mu
m$-to-11.3~\micron~ PAH features, independent of an AGN component. 
Table~\ref{regressionresults} includes the regression analysis for
the split samples. 

\subsubsection{The Effect of AGN}
\label{theeffectofagn}

The split-sample regression analysis for $L_{7.7}/L_{11.3}$
vs. $[OIII]/H\beta$ reveals that the correlation still holds for
the AGN+composite subsample with a similar slope to the full sample. While
there is a formal correlation for the star-forming subsample, the best-fit
slope is so shallow as to suggest that the correlation is not
meaningful. This suggests that there is a correlation between
the hardness of the radiation field originating from an AGN and 
the 7.7~\micron-to-11.3~\micron~ ratio. 

In Figure~\ref{BPThist} (left) we show our sample on the BPT diagram,
colour-coded as in Figure~\ref{BPT}. For better visual clarity,
symbol size is proportional to the ratio of long-to-short
wavelength PAH band luminosity (upper: $L_{7.7}/L_{6.2}$, 
lower: $L_{11.3}/L_{7.7}$), and hence to
average grain size and/or neutral PAH fraction. The histograms
(Fig.~\ref{BPThist}, right) show the distribution of PAH ratios for
the different BPT designations. The strongest trend is found for
$L_{7.7}/L_{11.3}$, for which the active galaxies show significantly
lower values than star-forming galaxies (with $<$0.1\% significance by
a KS test). 
More dominant AGN satisfying the Kewley designation have only
marginally lower $L_{7.7}/L_{11.3}$ than composite 
sources satisfying the Kauffmann designation (10\% significance). For 
$L_{6.2}/L_{7.7}$, the differences in the subsamples are
not highly significant; there is a 10\% significance for the difference
in this ratio between active and star-forming galaxies.

\begin{figure*}[ht!]
\centering
\includegraphics[angle=90,width=16cm]{f8.epsi}
\caption{{\it Left:} BPT plot for SSGSS sample.
Colour-coding is the same as in Figure~\ref{BPT}. 
Symbol size increases with decreasing short-to-long wavelength PAH ratios;
$L_{6.2}/L_{7.7}$ (upper) and $L_{7.7}/L_{11.3}$ (lower). The upper and
lower dotted lines in each plot represent the \citet{Kewley} and
\citet{Kauffmann03} divides respectively.
{\it Right:} Histogram of PAH ratios (upper: $L_{6.2}/L_{7.7}$, lower:
$L_{7.7}/L_{11.3}$) for star-forming galaxies, 
composite sources, and AGN.
\label{BPThist}}
\end{figure*}

Although there is only a tentative difference in the
$L_{7.7}/L_{11.3}$ distributions between the
BPT-designated full AGN and composite sources, these
galaxies have indistinguishable distributions of $D_n(4000)$
and H$\alpha$~EW by a KS test. Thus, any difference in PAH ratios is
likely due to AGN dominance and not to a link between AGN and star formation
properties. 

\begin{figure}[h!]
\vspace{1cm}
\hspace{-0.5cm}
\includegraphics*[width=8cm]{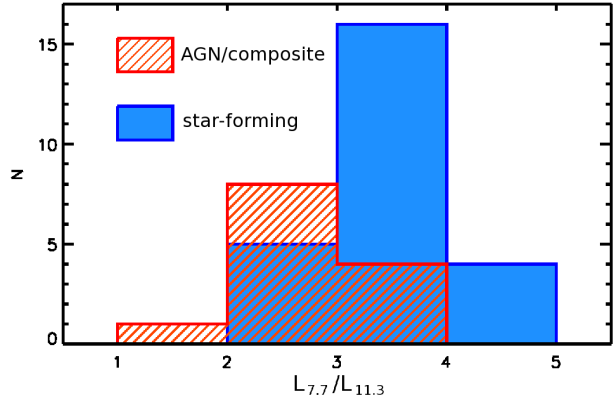}
\caption{
Histogram of $L_{7.7}/L_{11.3}$ for star-forming galaxies (blue) and
sources with any AGN component (red), only including sources with
$1.3 < D_n(4000) < 1.5$. A KS test indicates that samples are
different at a 2\% significance level.
\label{AGNvsNon_pahratiohist}}
\end{figure}

To better establish the independence
of this relationship to star formation we analyse a controlled subsample
restricted to a narrow age range, with $1.3 < D_n(4000) < 1.5$.
Figure~\ref{AGNvsNon_pahratiohist} shows the histogram of
$L_{7.7}/L_{11.3}$ for galaxies in this range, split between galaxies
with dominant AGN activity (either composite or full AGN by BPT designation),
versus galaxies with dominated by star formation.  
In this range, both subsamples have indistinguishable
distributions of both $D_n(4000)$ and H$\alpha$~EW. A KS
test of these PAH ratios shows that they are different, with a
significance level of 2\%. It is therefore likely that AGN play some role in
reducing the relative 7.7~\micron-to-11.3~\micron~ PAH emission, and that
the trend is not solely due to the link between AGN status and the age
of the stellar population.

\subsection{The Effect of Age, Metallicity and Hard Radiation Fields on PAH Molecules}
\label{hard}

\begin{figure*}[ht!]
\centering
\epsscale{.50}
\includegraphics[angle=90,width=16cm]{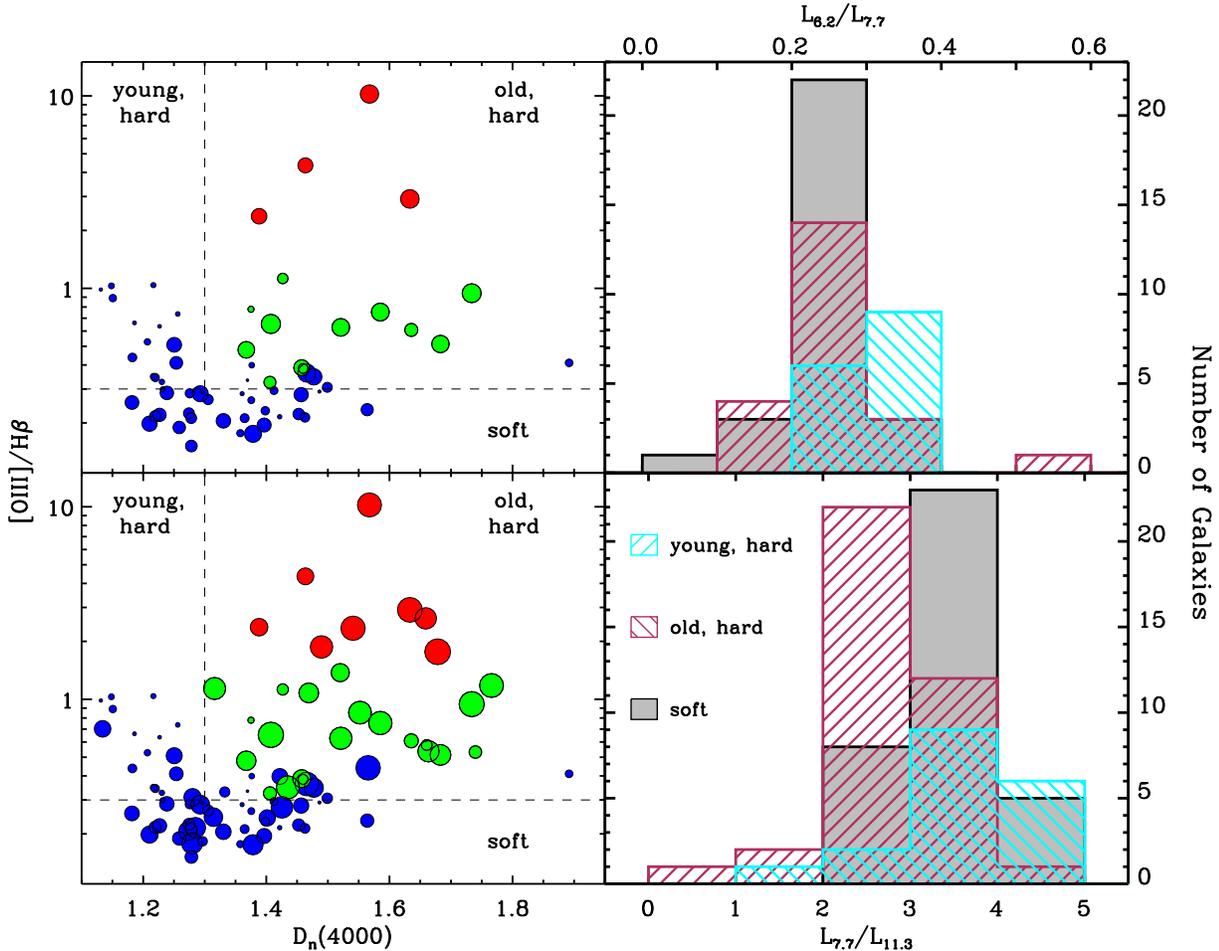}
\caption{{\it Left:} $[OIII]/H\beta$ vs. $D_n(4000)$. 
Colour-coding is the same as in Figure~\ref{BPT}. 
Symbol size increases with decreasing short-to-long wavelength PAH ratios;
$L_{6.2}/L_{7.7}$ (upper) and $L_{7.7}/L_{11.3}$ (lower).  
Dashed lines divide sample into the subpopulations used in the histogram.
{\it Right:} Histogram of PAH ratios (upper: $L_{6.2}/L_{7.7}$, lower:
$L_{7.7}/L_{11.3}$) for both old ($D_n(4000) > 1.3$) 
and young ($D_n(4000) < 1.3$) sources with hard radiation fields
($[OIII]/H\beta > 0.3$) and all sources with softer radiation fields
($[OIII]/H\beta < 0.3$).
\label{D4kBPThist}}
\end{figure*}

\begin{figure*}[ht!]
\includegraphics[width=13.5cm]{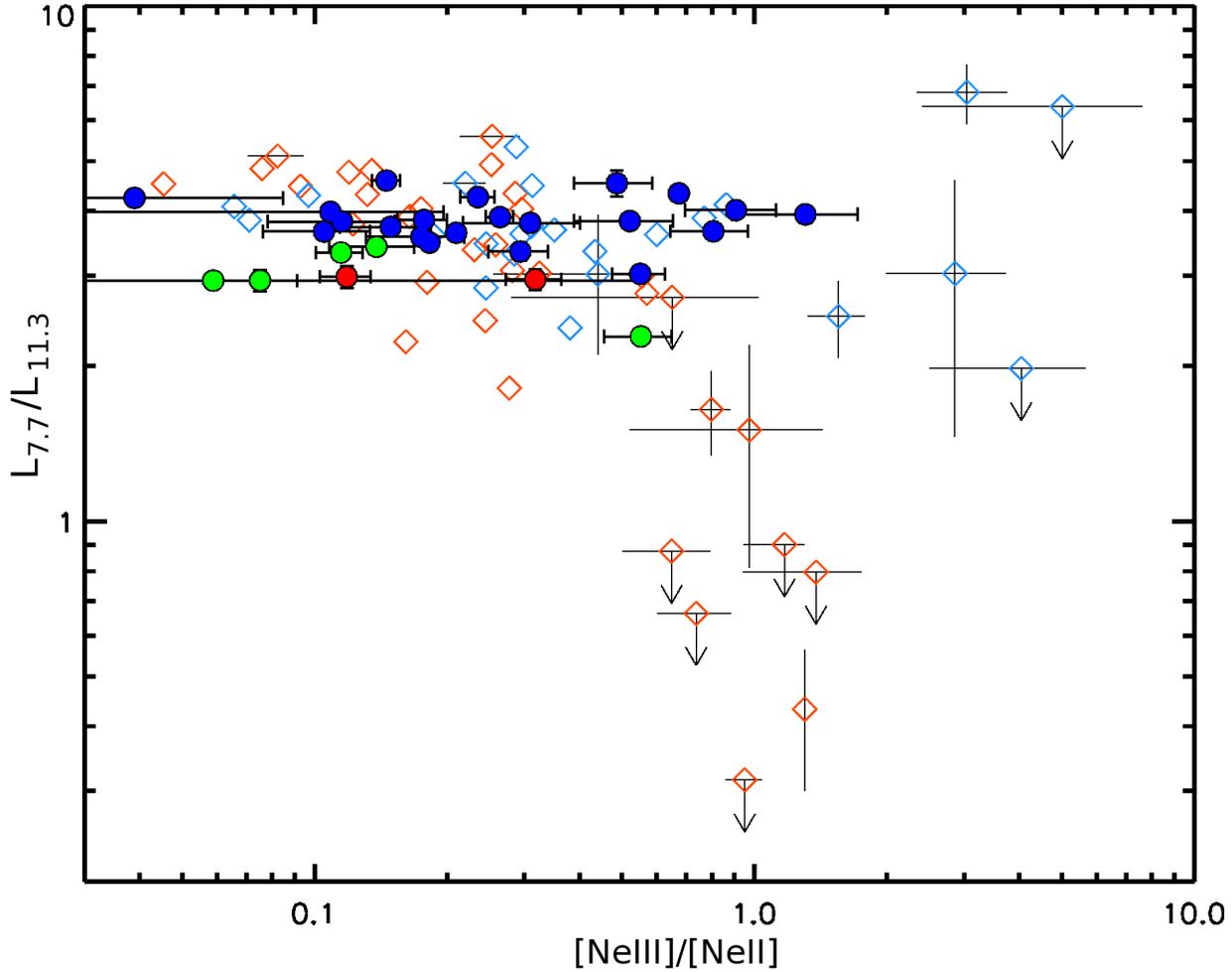}
\caption{$[NeIII]_{15.6~\mu m}/[NeII]_{12.8~\mu m}$ vs. $L_{7.7}/L_{11.3}$.
Circles are SSGSS galaxies, with colour-coding is the same as in
Figure~\ref{BPT}. Diamonds are SINGS galaxies from
\citet{Smith07}, with the red and blue points representing
AGN/composite sources and star-forming galaxies respectively.
\label{smithplot}}
\end{figure*}

In order to better separate the role of AGN and star formation, we
look at the relationship between hardness of the radiation field as
measured by $[OIII]/H\beta$ with stellar population age as measured
by $D_n(4000)$ (Fig.~\ref{D4kBPThist}). 
As a sensitive age diagnostic, $D_n(4000)$ is highly correlated with
[NII])/H$\alpha$, and so this plot is analogous to 
the BPT plot, but with a clearer separation between young and older
stellar populations. In this plot symbol size is again
proportional to long-to-short wavelength PAH ratio for the purpose of
visual clarity. Here we see that, for hard radiation
fields as measured by $[OIII]/H\beta$, there is a striking
difference in PAH ratio between younger and older
stellar populations. Younger stellar populations ($D_n(4000) < 1.3$)
with hard radiation fields ([OIII]/H$\beta > 0.3$) have stronger short
wavelength PAH bands than older galaxies with similar $[OIII]/H\beta$ 
(significance: 0.05\% for $L_{7.7}/L_{11.3}$, 3\% for $L_{6.2}/L_{7.7}$); 
they also have stronger short wavelength bands than galaxies of all
ages with softer radiation fields ([OIII]/H$\beta < 0.3$,
significance: 2\% for $L_{7.7}/L_{11.3}$, 1\% for $L_{6.2}/L_{7.7}$).
Older populations with hard ionizing fields (which are also
predominantly AGN-dominated) have stronger long wavelength
bands than the other populations  (significance: $<$0.01\% for 
$L_{7.7}/L_{11.3}$, 8\% for $L_{6.2}/L_{7.7}$).
The differences are illustrated in the histograms to the right of
Figure~\ref{D4kBPThist}.

Thus, it seems that the behaviour of PAH molecules in the presence of
a hard radiation field differs depending on the age of the stellar
population, and so it may depend on whether the radiation comes
predominately from starburst activity or an AGN. These trends may also
arise from a link between metallicity and both radiation field
hardness and stellar population. Below, we discuss the factors that
drive the evolution of 
short-to-long wavelength PAH ratios:

The youngest galaxies begin their evolution with low $D_n(4000)$, low
metallicity, a moderately hard radiation field resulting from
starburst activity, and high short-to-long wavelength PAH ratios.
The increase in metallicity with age is expected toallow 
the larger grains that dominate the longer wavelength feature in each
ratio to be assembled more efficiently, resulting in a drop in this
ratio as seen in Section~\ref{opticalcorrelation}.

Additionally, as age increases the starburst radiation field also
softens. This may result in a decrease in the ionization fraction of 
the PAH molecules, which, according to the \cite{DraineLi} models
(Fig.~\ref{DraineLi}), may contribute to the drop in
7.7~\micron-to-11.3~\micron~ ratio between the youngest galaxies in
this sample and older galaxies.

Even after these youngest, starburst stages, aging galaxies continue
to exhibit a decrease in short-to-long
wavelength PAH ratios, and in particular
in $L_{7.7}/L_{11.3}$ (see Fig.~\ref{PAHratiovsSF}). Some 
of these galaxies experience AGN activity, and so are subject to
radiation fields that are often 
much harder than in starburst galaxies. If the initial decrease in
PAH ratio with age is due to increasing neutral
fraction of PAH molecules, then AGN do not seem to reionize these
molecules; indeed, the trend of diminishing 
PAH ratio with galaxy age continues through an increasing incidence of AGN.

In Section~\ref{theeffectofagn} we saw that AGN seem to cause a decrease
in PAH ratio, particularly $L_{7.7}/L_{11.3}$, independently of star
formation diagnostics, and that $L_{7.7}/L_{11.3}$ also drops with AGN-sourced radiation hardness
(Fig.~\ref{PAHratiovsIonization}, lower left). This may be due to
preferential destruction of small PAH grains by X-rays and/or shocks
from the AGN. It may also be due to heating of PAH molecules by the
AGN, as ionization fraction decreases with increasing gas temperature
\citep{Tielens}. Thus, if the AGN is capable of
heating PAH to high temperatures in dense regions, compared to the
more diffuse heating by stars, then the net effect may be to reduce
the ionization fraction.


\subsubsection{[NeII] and [NeIII] Measurements}

This differing effect of an AGN- or starburst-sourced hard radiation
field on the PAH spectrum is
consistent with the result of S07, who plot
$L_{7.7}/L_{11.3}$ versus 
$[NeIII]_{15.6~\mu m}/[NeII]_{12.8~\mu m}$ for their sample. 
$[NeIII]/[NeII]$ provides a NIR measure of the
hardness of the radiation field. For a subsample of
galaxies with significant AGN components (Seyferts and LINERs), they
find a strong anti-correlation between the two ratios: the harder the
ionizing field in the AGN, the weaker the short wavelength PAH band.
However the hardness of the radiation field in star-forming galaxies --- radiation that
is primarily stellar in origin --- shows no correlation with PAH
ratio. \cite{Brandl} find a similar absense of correlation for 22
starburst nuclei over a similarly broad range of $[NeIII]/[NeII]$.

We have measured $[NeIII]/[NeII]$ for the subsample of 32 galaxies with
the IRS hi-res spectrograph. In Figure~\ref{smithplot} we compare our
results with S07.
Our results are consistent, however the SSGSS sample does not include $[NeIII]/[NeII]$
values as high as those found by S07, and so  we don't see the full
trend between these ratios. The likely reason for the difference in 
measured $[NeIII]/[NeII]$ is that for the local galaxies studied by
S07, the IRS slit samples a smaller, more central, and hence more
AGN-dominated region than for our  more distant galaxies.\\

\subsection{The 17~\micron~ Feature}

\begin{figure}[t!]
\vspace{2.5cm}
\hspace{-1cm}
\includegraphics[width=8.5cm]{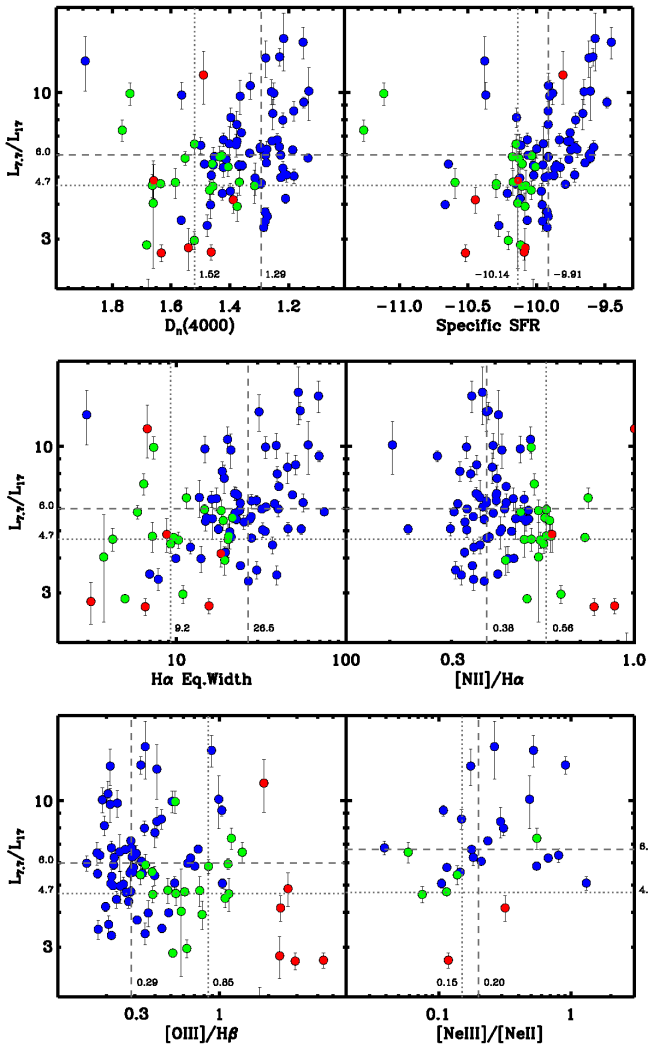}
\caption{
$L_{7.7}/L_{17}$ vs. SDSS optical diagnostics and $NeIII/NeII$.
Colour-coding is the same as in Figure~\ref{BPT}. 
\label{7on17vsOpt}}
\end{figure}

\begin{figure}[ht!]
\includegraphics*[width=8cm]{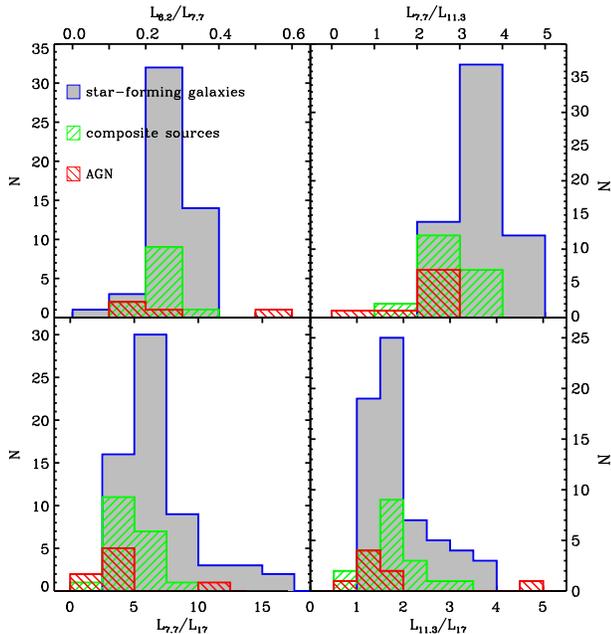}
\caption{PAH ratio by BPT designation, with grey for star-forming galaxies, green
  for composite sources and red for AGN. 
  {\it Top left}: $L_{6.2}/L_{7.7}$, 
  {\it top right}: $L_{7.7}/L_{11.3}$, 
  {\it bottom left}: $L_{6.2}/L_{17}$.
  {\it bottom right}: $L_{7.7}/L_{17}$.
\label{PAHhist}}
\end{figure}

The 17~\micron~ PAH band is well separated from the denser blueward
bands, and hence provides one of the cleanest measures of PAH
strength. This band may arise from C-C-C bending \citep{DraineLi2},
however the true vibrational mode or modes responsible for this emission
are not established. Nonetheless, it is expected that the dominant
contributing grain size will be larger than for lower wavelength bands. 
This band falls in the long-low IRS module, which has a significantly
larger slit width than both the short-low module and the SDSS
aperture, at 10\farcs5 vs. $\sim$3\farcs7 and 3\arcsec~ respectively. As a
result, the sampling of different regions of the galaxy introduces
unquantifiable uncertainties, and so we limit the analysis of this
feature to qualitative comparisons. 

In Figure~\ref{7on17vsOpt} we plot the ratio of the 7.7~\micron~ bands to 
the 17~\micron~ band versus SDSS optical properties. 
With respect to most optical diagnostics, $L_{7.7}/L_{17}$
displays similar correlations to $L_{7.7}/L_{11.3}$, with
similar dispersions. This ratio shows a significantly better
correlation with respect to SSFR, and a worse correlation with respect
to $[NII]/H\alpha$. As with $L_{7.7}/L_{17}$, no
correlation is observed with respect to $NeIII]/[NeII]$. 
Dividing the sample between BPT-designated AGN/composite sources and star-forming galaxies, the same
correlations are observed with all star formation diagnostics for
star-forming galaxies. The correlations are not significant
for other diagnostics or for the subsample of galaxies with AGN
components. 

In Figure ~\ref{PAHhist} we compare the histograms of the PAH ratios
$L_{6.2}/L_{7.7}$, $L_{7.7}/L_{11.3}$, 
$L_{7.7}/L_{17}$, and $L_{11.3}/L_{17}$
divided into subsamples of varying of AGN activity. As was seen in
Figure~\ref{BPThist}, $L_{6.2}/L_{7.7}$ reveals a slight
relative increase in the strength of the longer-wavelength PAH feature
for galaxies with an AGN component while
$L_{7.7}/L_{11.3}$ shows a much stronger increase. There
is also a significant increase in the strength of the 17~\micron~
feature relative to the 7.7~\micron~ feature
with increasing AGN incidence and power. There is an increase in the 
17~\micron~ feature relative to the 11.3~\micron~ feature between star-forming
galaxies and full AGN, but not between star-forming galaxies and
composite sources. 

This may indicate a threshold in AGN power needed
to destroy the larger PAH grains that dominate the 11.3~\micron~
feature; only full AGN are capable of destroying such grains with
enough efficiency to register a difference in the
$L_{11.3}/L_{17}$ ratio. Conversely, the average
molecule sizes that dominate both the 6.2 and 7.7~\micron~ features
are destroyed with similar efficiency even for composite sources,
resulting in little change in $L_{6.2}/L_{7.7}$ with AGN
status.

\begin{figure*}[ht!]
\centering
\includegraphics[angle=90,width=16cm]{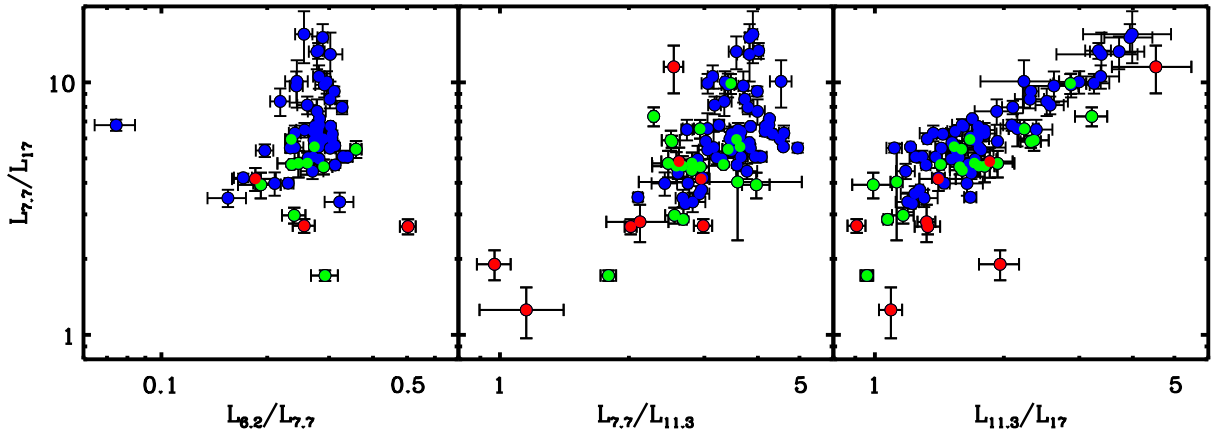}
\caption{
$L_{6.2}/L_{7.7}$ (left),
$L_{7.7}/L_{11.3}$ (middle), 
and $L_{11.3}/L_{17}$ (right) vs. $L_{7.7}/L_{17}$.
Colour-coding is the same as in Figure~\ref{BPT}. 
\label{17ratio}}
\end{figure*}

In Figure~\ref{17ratio} we study the 17~\micron~ feature in comparison
to shorter wavelength features. 
$L_{6.2}/L_{7.7}$ vs. $L_{7.7}/L_{17}$ 
shows a very narrow range of the former ratio over an order of
magnitude variation in the latter. 
$L_{7.7}/L_{17}$
correlates well with $L_{7.7}/L_{11.3}$. 
Interestingly, an even tighter correlation is observed between 
$L_{7.7}/L_{17}$ and $L_{11.3}/L_{17}$,
suggesting that the 7.7~\micron~ and 11.3~\micron~ features are more
closely tied than the  11.3~\micron~ and 17~\micron~ features. 

\section{CONCLUSIONS}
\label{con}

We have studied the relationships between a range of optical
diagnostics 
and the relative strengths of PAH emission bands for the SSGSS
sample of 92 {\em GALEX}-selected normal, star-forming galaxies observed
with the IRS lo-res and hi-res spectrographs. 
The observed PAH spectra are consistent with SINGS galaxies
studied by \cite{Smith07} and largely consistent with the 
galactic HII regions and galaxies studied by \cite{Galliano}, although
in the case of the latter, the different analysis methods appears to yield a
$\sim$20-50\% difference in the strength of the 6.2~\micron~ feature. 

Short-to-long wavelength
PAH feature ratios exhibit a number of trends with the SDSS optical
diagnostics of the galaxies' star formation histories, metallicities, and
radiation fields. The most striking of these is the correlation of the
7.7~\micron-to-11.3~\micron~ feature ratio with the star formation
diagnostics H$\alpha$~EW and $D_n(4000)$, and with the
emission line ratios $[NII]/H\alpha$ and
$[OIII]/H\beta$, and with metallicity. The correlation of this PAH
ratio with star formation diagnostics is independent
of the presence of an AGN component, indicating that the stellar
population plays an important role in determining the relative feature
strengths. This is consistent with an increase in production of
large-grain PAH molecules with increasing metallicity, and with
a correlation between the fraction
of PAH molecules that are ionized and the hardness of any starburst
radiation field. 

The presence of an AGN component, as determined by the galaxy's
location on the BPT diagram, is correlated with a reduction in the
ratio of the 7.7~\micron~ to 11.3 and 17~\micron~ features. For a
subsample with matched $D_n(4000)$ and H$\alpha$~EW, galaxies with
any AGN component (AGN or composite sources) have weaker relative
7.7~\micron~ emission than quiescent galaxies. This is consistent with
a picture in which smaller PAH grains are preferentially destroyed by
shocks and/or x-rays from the AGN.

\acknowledgments

We thank the anonymous referee for valuable comments that improved the
quality of this work.

This work is based on observations made with the 
{\it Spitzer Space Telescope}, which is operated by the Jet Propulsion 
Laboratory, California Institute of Technology under NASA contract
1407. 

GALEX is a NASA Small Explorer. We gratefully acknowledge NASA's
support for construction, operation, and science 
analysis for the GALEX mission, developed in cooperation with 
the Centre National d'Etudes Spatiale of France and the Korean 
Ministry of Science and Technology.

We thank the MPA/JHU collaboration for SDSS studies for making
their catalogs publicly available. 

This work utilized the PAHFIT IDL tool for decomposing IRS spectra, 
which J. D. Smith has generously made publically available \citep{Smith07}.

Funding for the SDSS has been provided by
the  Alfred P. Sloan Foundation, the Participating Institutions, NASA,
NSF, the U.S. Department of Energy, the Japanese Monbukagakusho, 
the Max Planck Society, and 
the Higher Education Funding Council for England. The SDSS 
Web site is http://www.sdss.org/. 
The SDSS is managed by the Astrophysical Research Consortium for the
Participating Institutions. The Participating Institutions are the
American Museum of Natural History, Astrophysical Institute Potsdam,
University of Basel, University of Cambridge, Case Western Reserve
University, University of Chicago, Drexel University, Fermilab, the
Institute for Advanced Study, the Japan Participation Group, Johns
Hopkins University, the Joint Institute for Nuclear Astrophysics, the
Kavli Institute for Particle Astrophysics and Cosmology, the Korean
Scientist Group, the Chinese Academy of Sciences (LAMOST), Los Alamos
National Laboratory, the Max-Planck-Institute for Astronomy (MPIA),
the Max-Planck-Institute for Astrophysics (MPA), New Mexico State
University, Ohio State University, University of Pittsburgh,
University of Portsmouth, Princeton University, the United States
Naval Observatory, and the University of Washington. 

{\it Facilities:} \facility{Spitzer (IRS), GALEX, SDSS}

\end{document}